\documentclass[11pt,a4paper]{article}

\usepackage{jheppub}

\title{A Massive S-duality in Four Dimensions}

\author[a]{Aybike \c{C}atal-\"{O}zer,}
\author[b]{Cemsinan Deliduman}
\author[c]{and E. Ula\c{s} Saka}

\affiliation[a]{Department of Mathematics, Faculty of Science and
Letters, \.{I}stanbul Technical University, Maslak 34469,
\.{I}stanbul, Turkey} \affiliation[b]{Department of Physics, Mimar
Sinan Fine Arts University, Bomonti 34380, \.{I}stanbul, Turkey}
\affiliation[c]{Department of Physics, \.{I}stanbul University,
Vezneciler 34134, \.{I}stanbul, Turkey}

\emailAdd{ozerayb@itu.edu.tr}
\emailAdd{cemsinan@msgsu.edu.tr}
\emailAdd{ulassaka@istanbul.edu.tr}

\abstract{We reduce the Type IIA supergravity theory with a
generalized Scherk-Schwarz ansatz that exploits the scaling
symmetry of the dilaton, the metric and the NS 2-form field. The
resulting theory is a new massive, gauged supergravity theory in
four dimensions with a massive 2-form field and a massive 1-form
field. We show that this theory is S-dual to a theory with a
massive vector field and a massive 2-form field, which are dual to
the massive 2-form and 1-form fields in the original theory,
respectively. The S-dual theory is shown to arise from a
Scherk-Schwarz reduction of the heterotic theory. Hence we
establish a massive, S-duality type relation between the IIA
theory and the heterotic theory in four dimensions. We also show
that the Lagrangian for the new four dimensional theory can be put
in the most general form of a $D=4, \ N=4$ gauged Lagrangian found
by Sch\"{o}n and Weidner, in which (part of) the $SL(2)$ group has
been gauged.}

\begin{document}
\maketitle

%%%%%%%%%%%%%%%%%%%%%%%%%%%%

\section{Introduction}

%%%%%%%%%%%%%%%%%%%%%%%%%%%%

String compactifications in the presence of fluxes has been an
important research area in recent years. Fluxes can be geometric
(like p-form or metric fluxes, see \cite{g} for a review) or non-geometric
\cite{stw,dpst}.
The importance of introducing flux into the compactification
scheme is that the lower dimensional theory is more realistic. The
resulting theory is gauged and massive with mass parameters
defining a scalar potential, which in turn gives rise to moduli
stabilization.

An important question is the faith of string dualities, when
fluxes are introduced. One of the oldest work, which explored this
question is that of Kaloper and Myers \cite{km}, who considered flux
compactifications of the heterotic string on the $d$-dimensional
torus $T^d$. They showed that the perturbative $O(d,d+16)$ duality
symmetry  is still a symmetry of the resulting gauged, massive
supergravity, provided that the mass parameters also transform
under the duality group. On the other hand, flux compactifications
of Type II theories on Calabi-Yau manifolds were studied in \cite{lm}.
In the papers \cite{lm,glmw} and later in
\cite{glw} it was established that the mirror symmetry
between IIA and IIB theories is still valid in the presence of
fluxes. The U-duality symmetry of M-theory compactifications with
flux was explored in \cite{hre}.

Although much has been understood about the perturbative duality
symmetries in flux compactifications, less is known about the
non-perturbative ones. For example, the six-dimensional theory
obtained from the compactification of IIA theory on $K3$ manifold
is known to be S-dual to heterotic string theory compactified on
$T^4$ \cite{ht,ew,as,hs,Behrndt:1995si}. Similarly, heterotic
string theory compactified on $K3 \times T^2$ to four dimensions
is S-dual to Type IIA theory compactified on a certain Calabi-Yau
manifold \cite{js,jss1,jss2,as2,d}. It is natural to ask whether
these duality relations continue to hold when fluxes are turned
on. The main aim of the present paper is to contribute towards
answering this question.

For the S-duality in four dimensions there were early attempts \cite{ckkl,lm2},
which identified the duals of some fluxes that can be
introduced in the heterotic compactification. In more recent work
\cite{ablm}, it was suggested that in order to find the
duals of all heterotic fluxes, the IIA theory has to be lifted to
M-theory.

For the six-dimensional S-duality symmetry with fluxes,  earlier
work was done in \cite{bbrs}, where it is argued that the duality
does not hold at the level of the action, when fluxes are
introduced. However, they were able to establish a six dimensional
massive S-duality by performing a Scherk-Schwarz reduction of
seven dimensional IIA theory, obtained by a K3 compactification of
M-theory, and the heterotic theory compactified on $T^3$. On the
other hand, in \cite{hls} flux compactification of massive IIA
theory was performed. The resulting theory was shown to possess
the perturbative $O(4,20)$ symmetry. However, the S-duality
symmetry which is to map the theory to heterotic theory could not
be restored. The main problem is identified to be due to the fact
that on the IIA side it is the NS-NS 2-form field which acquires
mass, whereas on the heterotic side vector fields get massive. It
might be possible to resolve this problem in four dimensions,
since a four dimensional massive 2-form field has the same number
of degrees of freedom as a massive vector field in four
dimensions. So one can consider to perform a further $T^2$
reduction of both theories and seek the desired massive S-duality
in four dimensions \cite{singh}. In the massless case, it is well
known that both theories have an $O(6,22) \times SL(2)$ symmetry
in four dimensions. On the IIA side, one has the perturbative
$O(4,20)$ symmetry due to K3 compactification, combined with the
$SL(2) \times SL(2)$ symmetry of the $T^2$ compactification. Under
S-duality, the $SL(2)$ symmetry associated with the torus
compactification gets mapped to the self-duality of the heterotic
theory on $T^6$, whereas $O(6,22)$ is the T-duality symmetry of
the heterotic theory associated with the compactification manifold
$T^6$. When  fluxes are introduced, the $O(4,20) \times SL(2)
\times SL(2)$ symmetry of the IIA theory was shown to remain as
the symmetry group, provided that the mass parameters also
transform under this duality group
\cite{bbrs,hls,singh,Janssen:2001hy}. On the heterotic side,
although the $O(6,22)$ part is still a symmetry as was shown by
Kaloper and Myers \cite{km}, the self-duality $SL(2)$ is
problematic. Recall that in four dimensional heterotic theory, one
dualizes the 2-form field coming from the reduction of the NS-NS
2-form field to a scalar, which then forms an $SL(2)$ doublet
along with the dilaton. This dualization can no longer be
performed in the presence of fluxes, as fluxes imply non-abelian
gauge couplings for the 2-form field. However, if the massive
S-duality is to hold in 4 dimensions, one expects that it maps the
massive IIA theory (for which the $SL(2)$ is still a symmetry) to
a massive heterotic theory, which still possesses the self-duality
 symmetry. One way to approach this problem is to start with the
general $SL(2)$-gauged supergravity in 4 dimensions, seek for a
string theory origin and see if this teaches us something about
the (possible)  massive S-duality between the IIA and the
heterotic theory. This is the way we approach the problem of
massive S-dualities in this paper.

The most general $O(6,22) \times SL(2)$ gauged supergravity was
constructed by Sch\"{o}n and Weidner in \cite{sw}. The
string/M-theory origin of the most general $SL(2)$ gauging is
still not known. However, for certain types of gaugings, namely
for those which correspond to scalings and shifts of axion and
dilaton in four dimensions, a higher dimensional origin was found
by Derendinger et al. \cite{dpp}. They showed that the dimensional
reduction of the ten dimensional heterotic string (to be more
precise, the dimensional reduction of  the NS sector of the
heterotic string, without the Yang-Mills vectors) with a
generalized Scherk-Schwarz ansatz gives  in four dimensions, after
certain dualizations, the Sch\"{o}n-Weidner Lagrangian with
non-zero $SL(2)$ gaugings. In this paper, we utilize a similar
Scherk-Schwarz ansatz for the reduction of the six dimensional
type IIA theory (Such reductions were also considered by
\cite{zwirner1, zwirner2}). We show that the resulting massive
theory is S-dual to heterotic string theory
 reduced with the
Scherk-Schwarz twist of Derendinger et al. \cite{dpp}.
Although we work with a restricted class of fluxes,
our work is interesting, because it gives an explicit
demonstration of how the duality between massive 2-forms and
massive 1-forms work in the context of string theory. As a
by-product we show that the inclusion of the Yang-Mills vectors to
the heterotic string theory does not change the results of
Derendinger et al. \cite{dpp}. The resulting gauged supergravity is still of
the Sch\"{o}n-Weidner type, characterized by the same
embedding tensor.

The plan of our paper is as follows. In section 2, we introduce
the aforementioned Scherk-Schwarz twist and perform the dimensional
reduction of the six-dimensional Type IIA theory. In section 3 we
dualize the resulting theory and show that the dual theory can be
obtained from a dimensional reduction of the heterotic theory.
In section 4, we discuss in more detail how the duality between
the massive 2-form fields and the massive 1-form fields work. In
section 5 we show that the dual massive theory can be put in the form
of  Sch\"{o}n-Weidner Lagrangian, and the gaugings are described by
the same tensor as the one in Derendinger et al, although we also
include the Yang-Mills vectors. We discuss our results in section 6.

%%%%%%%%%%%%%%%%%%%%%%%%%%%%

\section{Twisted Reduction of Type IIA theory from 6 to 4 Dimensions}

%%%%%%%%%%%%%%%%%%%%%%%%%%%%

In this section we perform a dimensional reduction of the
six-dimensional Type IIA theory to four dimensions on a two-torus
with a certain Scherk--Schwarz twist \cite{ss1,ss2}. The
six-dimensional Type IIA Lagrangian is obtained by a standard
Kaluza--Klein reduction of the ten-dimensional Type IIA
supergravity on K3 \cite{ht,ew,as,hs}. The field content of the
ten-dimensional Type IIA supergravity consists of a dilaton, a
two-form Kalb-Ramond field, and a one- and a three-form
Ramond-Ramond fields. The bosonic part of the six-dimensional Type
IIA Lagrangian, given as
\begin{eqnarray}
\mathcal{L}^{IIA}_{6} &=& e^{-\phi} \left( R * 1 - d\phi \wedge * d\phi +
\dfrac{1}{4} d\widetilde{M}_{IJ} \wedge * d\widetilde{M}^{IJ}   -\dfrac{1}{2}H_{(3)} \wedge
* H_{(3)} -\dfrac{1}{2} e^{\phi} \widetilde{M}_{IJ} F_{(2)}^I \wedge * F_{(2)}^J \right) \nonumber \\
& &  - \dfrac{1}{2} L_{IJ} B_{(2)} \wedge F_{(2)}^I \wedge F_{(2)}^J , \label{IIA-6}
\end{eqnarray}
is $O(4,20)$-invariant and the full theory has $\mathcal{N}=2$
supersymmetry in six dimensions. Here $\widetilde{M}_{IJ}$ with
$I=1, \ldots, 24$ is the scalar matrix that takes values in
$O(4,20)/O(4) \times O(20)$ coset space; $L_{IJ}$ is the invariant
metric of $O(4, 20)$; $H_{(3)} = dB_{(2)}$; and finally $F_{(2)}^I
= dA_{(1)}^I$, where $A_{(1)}^I$ is the $O(4,20)$ vector, $
A_{(1)}^I = ( A_{(1)}^a , B_{(1)  a} , A_{(1)}^A )$, with
$a=1,\dots , 4$, and $A=1,\ldots , 16$.\footnote{How $A_{(1)}^I$ is related to
ten-dimensional vector fields of Type IIA theory is explained, for
example, in \cite{res}.} We reduce this Lagrangian on a 2-torus $T^2$ with
the following twisted ansatz
\begin{eqnarray}
\phi(x, y) & = & \widetilde{\phi}(x) - 2 \lambda_m y^m,  \nonumber \\
G(x,y) & = & e^{-\lambda_m y^m}
\left( \widetilde{G} + \widetilde{G}_{mn}\eta^m \otimes \eta^n \right) , \nonumber \\
B_{(2)}(x,y) & = & e^{-\lambda_m y^m} \left(
\widetilde{B}_{(2)} + \widetilde{B}_{(1)m} \wedge \eta^m
+ \frac12 \widetilde{B}_{(0)mn}\eta^m \wedge \eta^n \right) , \nonumber \\
%H_{(3)}(x,y) & = & e^{-\lambda_m y^m} \left(
%\widetilde{H}_{(3)} + \widetilde{H}_{(2)m} \wedge \eta^m
%+ \frac12 \widetilde{H}_{(1)mn}\eta^m \wedge \eta^n \right) , \nonumber \\
A_{(1)}^I(x,y) & = & e^{1/2 \lambda_m y^m} \left(
\widetilde{A}_{(1)}^I + \widetilde{A}_{(0)m}^I \wedge \eta^m
\right), \label{SS}
%F_{(2)}^I (x,y) & = & e^{1/2 \lambda_m y^m} \left(
%\widetilde{F}_{(2)}^I + \widetilde{F}_{(1)m}^I \wedge \eta^m +
%\frac12 \widetilde{F}_{(0)mn}^I \eta^m \wedge \eta^n \right).
\end{eqnarray}
Here $y^m$ with $m=1,2$ are the coordinates on $T^2$, the
parameters $\lambda_m$ are arbitrary real numbers, $\eta^m = dy^m
+ \widetilde{\mathcal{A}}_{(1)}^m$, and
$\widetilde{\mathcal{A}}_{(1)}^m$ is the graviphoton of the
reduction. In this notation $\Omega_{(p)}$ is a p-form in six
dimensions and $\widetilde{\Omega}_{(p)}$ is a p-form in four
dimensions. This type of reduction is different from the
Kaluza--Klein reduction in the sense that one takes into account
not just the zeroth order term, but also the higher order terms in
the harmonic expansion of fields on the compactification manifold,
here $T^2$. However, dependence of the fields on the coordinates
of the internal manifold cannot be arbitrary. The reduced
Lagrangian should be  independent of the coordinates of the
compactification manifold. To attain to this requirement one has
to choose the Scherk-Schwarz reduction ansatz according to some
symmetry of the theory \cite{bde,kkm,h,hc,haybike,re,res}. The
reduction ansatz above is dictated by the $SL(2,R)$ scaling
symmetry of the two-torus:
\begin{equation}
\label{sym} \phi \rightarrow \phi - 2 \lambda ,\quad G \rightarrow
e^{-\lambda} G(x),\quad B_{(2)} \rightarrow e^{-\lambda}
B_{(2)}(x)
\end{equation}
This symmetry ensures that the ansatz (\ref{SS}) yields a
consistent reduction. We first reduce the Einstein-Hilbert part
together with the dilaton kinetic term of the six-dimensional Type
IIA Lagrangian (\ref{IIA-6}). We perform the reduction of the
Ricci scalar by expressing it in the so called Palatini form
\cite{o}. By utilizing a standard  ansatz for the vielbein we
calculate the non-vanishing components of the anholonomy
coefficients \cite{o} and the corresponding spin connection
components, in terms of which the Palatini form is given. Then the
usual reduction of the metric is performed. To absorb the volume
form of the compactification manifold it is also necessary to
shift the dilaton and define the four-dimensional dilaton as
$\widetilde{\phi}=\phi - \frac12 \log \det \widetilde{G}_{mn}$,
where $\widetilde{G}_{mn}$ is a symmetric 2 by 2 metric on $T^2$.
In order to write the action in the Einstein frame, we also
perform a conformal rescaling of the four-dimensional metric,
$\widetilde{G}_{\mu\nu} \rightarrow \frac{e^{\widetilde{\phi}}}2
\widetilde{G}_{\mu\nu}$, with $\mu,\nu=0,1,2,3$, and also a final
rescaling $\widetilde{\phi} \rightarrow 2\widetilde{\phi}$ of the
dilaton. The reduced form of the first two terms of (\ref{IIA-6})
in the Einstein frame are then found to be
\begin{eqnarray}
\mathcal{L}^{IIA}_{4,\ gravity} &=&  \frac12 \widetilde{R} *1
+\frac18 D\widetilde{G}_{mn} \wedge *D\widetilde{G}^{mn} \label{IIA-4g} \\
& & - D\widetilde{\phi}\wedge * D\widetilde{\phi} -\frac14
e^{-2\widetilde{\phi}}\widetilde{G}_{mn}\mathcal{F}_{(2)}^m \wedge
* \mathcal{F}_{(2)}^{n} -\frac12 e^{2\widetilde{\phi}} \lambda_m
\widetilde{G}^{mn} \lambda_n , \nonumber
\end{eqnarray}
where $\mathcal{F}^m_{(2)}=d \mathcal{A}_{(1)}^m$ is the field
strength of the graviphoton, $D\widetilde{\phi}=d
\widetilde{\phi}-\frac12 \lambda_k \mathcal{A}_{(1)}^k$.

 We now
insert the reduction ansatz, (\ref{SS}), into the NS-NS part of
the Lagrangian (\ref{IIA-6}) and obtain in four dimensions an
effective theory with the Lagrangian,
\begin{eqnarray}
\mathcal{L}^{IIA}_{4,\ NS-NS} &=& -e^{-\widetilde{\phi}} \left[ \dfrac{1}{2}
\widetilde{H}_{(3)} \wedge * \widetilde{H}_{(3)} +\dfrac{1}{2}
\widetilde{H}_{(2)m}\widetilde{G}^{mn} \wedge * \widetilde{H}_{(2)n} \right. \nonumber \\
&& +\dfrac{1}{2} \widetilde{H}_{(1)mn} \widetilde{G}^{mp}
\widetilde{G}^{nq} \wedge * \widetilde{H}_{(1)pq}
+\dfrac{1}{2} \widetilde{M}_{IJ} \widetilde{F}_{(2)}^I
\wedge * \widetilde{F}_{(2)}^J \label{IIA-4} \\
&& + \left. \dfrac{1}{2} \widetilde{M}_{IJ} \widetilde{F}_{(1)m}^I \widetilde{G}^{mn} \wedge *
\widetilde{F}_{(1)n}^J +\dfrac{1}{2} \widetilde{M}_{IJ}
\widetilde{F}_{(0)mn}^I \widetilde{G}^{mp}
\widetilde{G}^{nq} \wedge * \widetilde{F}_{(0)pq}^J \right] + \mathcal{L}_{CS}. \nonumber
\end{eqnarray}
where $\mathcal{L}_{CS}$ contains the Chern-Simons terms of the Lagrangian
$\mathcal{L}^{IIA}_{4}$:\footnote{To agree with the conventions of \cite{dpp} we take
$\psi_{[m}\chi_{n]}=\frac12 (\psi_{m}\chi_{n} - \psi_{n}\chi_{m})$.}
\begin{eqnarray}
\mathcal{L}_{CS} & = & - L_{IJ} \epsilon^{mn} \widetilde{B}_{(2)} \wedge
\widetilde{F}_{(1)[m}^I \wedge \widetilde{F}_{(1)n]}^J
- \dfrac{1}{2} L_{IJ} \epsilon^{mn} \widetilde{B}_{(2)} \wedge
\widetilde{F}_{(2)}^I \wedge \widetilde{F}_{(0)mn}^J \nonumber \\
&& -2 L_{IJ} \epsilon^{mn} \widetilde{B}_{(1)[m} \wedge \widetilde{F}_{(2)}^I
\wedge \widetilde{F}_{(1)n]}^J - \dfrac{1}{4} L_{IJ} \epsilon^{mn} \widetilde{B}_{(0)mn}
\wedge \widetilde{F}_{(2)}^I \wedge \widetilde{F}_{(2)}^J. \label{IIA-4CS}
\end{eqnarray}

\noindent The four dimensional fields that appear in the above
Lagrangian are obtained from
\begin{eqnarray}
F_{(2)}^I (x,y) & = & e^{1/2 \lambda_m y^m} \left(
\widetilde{F}_{(2)}^I + \widetilde{F}_{(1)m}^I \wedge \eta^m +
\frac12 \widetilde{F}_{(0)mn}^I \eta^m \wedge \eta^n \right), \nonumber \\
H_{(3)}(x,y) & = & e^{-\lambda_m y^m} \left( \widetilde{H}_{(3)} +
\widetilde{H}_{(2)m} \wedge \eta^m + \frac12
\widetilde{H}_{(1)mn}\eta^m \wedge \eta^n \right). \label{SS2}
\end{eqnarray}
Their explicit forms are
\begin{eqnarray}
\widetilde{F}_{(2)}^I & = & d\widetilde{A}_{(1)}^I + \widetilde{A}_{(0)m}^I {\mathcal{F}}_{(2)}^m +
\dfrac{1}{2} \lambda_m \widetilde{A}_{(1)}^I \wedge {\mathcal{A}}_{(1)}^m
\equiv D\widetilde{A}_{(1)}^I + \widetilde{A}_{(0)m}^I {\mathcal{F}}_{(2)}^m \ , \nonumber \\
\widetilde{F}_{(1)m}^I & = & d\widetilde{A}_{(0)m}^I - \dfrac{1}{2} \lambda_r
{\mathcal{A}}_{(1)}^r \widetilde{A}_{(0)m}^I - \dfrac{1}{2} \lambda_m \widetilde{A}_{(1)}^I \equiv
D\widetilde{A}_{(0)}^I - \dfrac{1}{2} \lambda_m \widetilde{A}_{(1)}^I\ , \nonumber \\
\widetilde{F}_{(0)mn}^I & = & \lambda_{[m} \widetilde{A}_{(0)n]}^I
\ , \label{IIA-F}
\end{eqnarray}

\noindent and
\begin{eqnarray}
\widetilde{H}_{(3)} & = & d\widetilde{B}_{(2)} + \lambda_r \widetilde{B}_{(2)} {\mathcal{A}}_{(1)}^r - \widetilde{B}_{(1)m} {\mathcal{F}}_{(2)}^m
\equiv \widetilde{D}\widetilde{B}_{(2)} - \widetilde{B}_{(1)m} {\mathcal{F}}_{(2)}^m \ ,  \nonumber \\
\widetilde{H}_{(2)m} & = & d\widetilde{B}_{(1)m} - \lambda_r
\widetilde{B}_{(1)m} {\mathcal{A}}_{(1)}^r - \lambda_m
\widetilde{B}_{(2)} - \widetilde{B}_{(0)mn} {\mathcal{F}}_{(2)}^n
\nonumber \\
& \equiv & \widetilde{D}\widetilde{B}_{(1)m} - \lambda_m
\widetilde{B}_{(2)}
- \widetilde{B}_{(0)mn} {\mathcal{F}}_{(2)}^n \ , \nonumber \\
\widetilde{H}_{(1)mn} & = & d\widetilde{B}_{(0)mn}
+ \lambda_r \widetilde{B}_{(0)mn} {\mathcal{A}}_{(1)}^r
+2\lambda_{[m}\widetilde{B}_{(1)n]} \equiv \widetilde{D}\widetilde{B}_{(0)mn}\ . \label{IIA-H}
\end{eqnarray}
The scalar matrix $\widetilde{M}_{IJ}$ is $O(4,20)/O(4) \times O(20)$ valued and it is given in terms
of the geometric moduli on K3 and components of the $B-$field wrapping the harmonic cycles of K3.
We do not need its explicit form here, which can be found in many sources, e.g. in \cite{hre}.

The twisted reduction ansatz we employ here exploits the scaling
symmetry of the dilaton, metric, the NS 2-form field and the
vectors $A_{(1)}^I$. Therefore, these fields are charged under the
gauge symmetry of the lower dimensional theory and their
derivatives become covariant derivatives as above. The gauge field
is the graviphoton  $\mathcal{A}_{(1)}^m$, which is the vector
field that comes from the reduction of the metric.

%%%%%%%%%%%%%%%%%%%%%%%%%%%%

\section{The Heterotic S-dual in 4 Dimensions}

%%%%%%%%%%%%%%%%%%%%%%%%%%%%

 In this section, we
dualize the 4d IIA theory (\ref{IIA-4g}, \ref{IIA-4},
\ref{IIA-4CS}) that we obtained in the previous section and show
that the resulting massive/gauged theory can be obtained from a
twisted reduction of the heterotic string theory. The dualization
is nontrivial as the fields that will be dualized, namely
$B_{(2)}, B_{(1)m}$ and $B_{(0)mn}$ appear through not only their
field strengths, but also through their bare potentials. We
overcome this difficulty in two steps. Firstly, we rewrite the
Chern-Simons (CS) term (\ref{IIA-4CS}) by adding total derivative
terms that will not alter the field equations, such that the
resulting CS term involves only the field strengths of the
relevant fields. Secondly, we add to the Lagrangian several
Lagrange multiplier terms, which couple the (field strengths of)
the fields that will be dualized to not only the field strengths
but also to the bare potentials of the ``dual-to-be" fields. The
duality is of the S-duality type, because under this duality the
dilaton, whose expectation value determines the string coupling
constant, changes sign.

The relevant Lagrange multiplier terms are:

\begin{eqnarray}
\left( d\widehat{B}_{(2)} - \lambda_r \widehat{B}_{(2)} {\mathcal{A}}_{(1)}^r
- \widehat{B}_{(1)m} {\mathcal{F}}_{(2)}^m  \right)
&\wedge & \widetilde{H}_{(1)mn}\epsilon^{mn} \nonumber \\
+ \left( d\widehat{B}_{(1)m} + \lambda_r \widehat{B}_{(1)m} {\mathcal{A}}_{(1)}^r
+ \lambda_m \widehat{B}_{(2)} - \widehat{B}_{(0)mr} {\mathcal{F}}_{(2)}^r \right)
&\wedge & \widetilde{H}_{(2)n} \epsilon^{mn} \nonumber \\
+ \left( d\widehat{B}_{(0)mn} - \lambda_r \widehat{B}_{(0)mn} {\mathcal{A}}_{(1)}^r
- 2\lambda_{[m}\widehat{B}_{(1)n]} \right)
&\wedge & \widetilde{H}_{(3)} \epsilon^{mn} . \label{Lmult}
\end{eqnarray}
Variation of the Lagrangian with respect to the fields $\widehat{B}_{(2)}$,
$\widehat{B}_{(1)m}$ and $\widehat{B}_{(0)mn}$
impose three different identities that the field strengths $\widetilde{H}_{(3)}$,
$\widetilde{H}_{(2)m}$ and $\widetilde{H}_{(1)mn}$ should obey. These identities are respectively,
\begin{eqnarray}
-\widetilde{D}\widetilde{H}_{(1)mn} - 2\lambda_{[m}\widetilde{H}_{(1)n]} & = & 0, \nonumber \\
\widetilde{D}\widetilde{H}_{(2)n} + \lambda_{n}\widetilde{H}_{(3)}
- \mathcal{F}^m \widetilde{H}_{(1)mn} & = &  0, \nonumber \\
\widetilde{D}\widetilde{H}_{(3)} + \mathcal{F}^m \widetilde{H}_{(2)m} & = & 0 .
\end{eqnarray}
These are precisely the Bianchi identities that should be
satisfied by $\widetilde{H}_{(3)}, \widetilde{H}_{(2)n}  $ and
$\widetilde{H}_{(1)mn} $, as can be checked straightforwardly from
(\ref{IIA-H}).

To perform the variation of the Lagrangian with respect to the field strengths
$\widetilde{H}_{(3)}$, $\widetilde{H}_{(2)m}$ and $\widetilde{H}_{(1)mn}$ we first need to write the
Chern-Simons part of the four-dimensional Type IIA Lagrangian (\ref{IIA-4CS}) in terms of these
fields. After some work we find that $\mathcal{L}_{CS}$ can be written as
\begin{eqnarray}
\mathcal{L}_{CS} & = & \dfrac{1}{4} L_{IJ} \epsilon^{mn} \widetilde{H}_{(3)}
\wedge \widetilde{A}_{(1)}^I \wedge \widetilde{F}_{(0)mn}^J
- L_{IJ} \epsilon^{mn} \widetilde{H}_{(3)} \wedge
\widetilde{A}_{(0)[m}^I \wedge \widetilde{F}_{(1)n]}^J \nonumber \\
&& + L_{IJ} \epsilon^{mn} \widetilde{H}_{(2)[m} \wedge \widetilde{A}_{(1)}^I
\wedge \widetilde{F}_{(1)n]}^J + L_{IJ} \epsilon^{mn} \widetilde{H}_{(2)[m}
\wedge \widetilde{A}_{(0)n]}^I \wedge \widetilde{F}_{(2)}^J \nonumber \\
&& +\frac14 L_{IJ} \epsilon^{mn} \widetilde{H}_{(1)mn} \wedge
\widetilde{A}_{(1)}^I \wedge \widetilde{F}_{(2)}^J, \label{L-CS}
\end{eqnarray}
together with some complicated total derivative terms, which will not contribute to any equation
obtained through variation of the action.

The variation of the Lagrangian (sum of eqs. (\ref{IIA-4}), (\ref{L-CS}) and (\ref{Lmult}))
with respect to the field strengths
$\widetilde{H}_{(3)}$, $\widetilde{H}_{(2)m}$ and $\widetilde{H}_{(1)mn}$ gives, respectively,
\begin{eqnarray}
e^{-\widetilde{\phi}}\epsilon^{mn} * \widetilde{H}_{(1)mn} & = &
\widehat{D}\widehat{B}_{(2)} - \widehat{B}_{(1)m} {\mathcal{F}}^m_{(2)}
- \dfrac{1}{2} L_{IJ} \widehat{A}_{(1)}^I \wedge \widehat{F}_{(2)}^J
\equiv \widehat{H}_{(3)} \label{4d-duality} \\
e^{-\widetilde{\phi}} \epsilon_m^{\ \ n} * \widetilde{H}_{(2)n} & = &
\widehat{D}\widehat{B}_{(1)m} + \lambda_m \widehat{B}_{(2)}
- \widehat{B}_{(0)mn} {\mathcal{F}}^n_{(2)}
-\frac12 L_{IJ} \widehat{A}_{(0)m}^I \widehat{F}_{(2)}^J
-\frac12 L_{IJ} \widehat{A}_{(1)}^I \widehat{F}_{(1)m}^J \nonumber \\
&\equiv & \widehat{H}_{(2)m} \nonumber \\
e^{-\widetilde{\phi}}\epsilon_{mn} * \widetilde{H}_{(3)} & = &
\widehat{D}\widehat{B}_{(0)mn} - \frac12 L_{IJ} \widehat{A}_{(1)}^I \widehat{F}_{(0)mn}^J
+ L_{IJ} \widehat{A}_{(0)[m}^I \widehat{F}_{(1)n]}^J
\equiv \widehat{H}_{(1)mn} ,  \nonumber
\end{eqnarray}
where the covariant derivatives  are defined as
\begin{eqnarray}
\widehat{D}\widehat{B}_{(2)} &=& d\widehat{B}_{(2)}
- \lambda_r \widehat{B}_{(2)} {\mathcal{A}}_{(1)}^r \nonumber \\
\widehat{D}\widehat{B}_{(1)m} &=& d\widehat{B}_{(1)m}
+ \lambda_r \widehat{B}_{(1)m} {\mathcal{A}}_{(1)}^r \nonumber \\
\widehat{D}\widehat{B}_{(0)mn} &=& d\widehat{B}_{(0)mn} -
\lambda_r \widehat{B}_{(0)mn} {\mathcal{A}}_{(1)}^r -
2\lambda_{[m}\widehat{B}_{(1)n]}\ . \label{covdev}
\end{eqnarray}

Next we make the identifications
\begin{equation}
\label{MandA}
\widetilde{M}_{IJ} \rightarrow \widehat{M}_{IJ}\ , \quad
\widetilde{A}^I_{(1)} \rightarrow \widehat{A}^I_{(1)} .
\end{equation}
The first identification here is understood as such that the
scalar matrix of heterotic theory is constructed in terms of the
geometric moduli of $T^4$ and the expectation value of the B-field
on $T^4$. However, its form is the same as that of
$\widetilde{M}_{IJ}$, for it still should be $O(4,20)/O(4) \times
O(20)$ valued. This scalar matrix is given as
\begin{equation}
\widehat{M}^{IJ} = \left(\begin{array}{ccc}
  \widehat{G} + \widehat{C}^T \widehat{G}^{-1} \widehat{C}
  + \widehat{A}^{T} \widehat{A}
  & -\widehat{C}^T \widehat{G}^{-1}
  & \widehat{C}^T \widehat{G}^{-1}L \widehat{A} + \widehat{A}^{T} \\
  -\widehat{G}^{-1} \widehat{C} & \widehat{G}^{-1} & -\widehat{G}^{-1} L \widehat{A} \\
  \widehat{A}^T L \widehat{G}^{-1} \widehat{C} + \widehat{A} &
  -\widehat{A}^T L \widehat{G}^{-1} & 1 + \widehat{A}^T L \widehat{G}^{-1} L \widehat{A} \\
\end{array}\right). \label{Het-M}
\end{equation}

\noindent Here $\widehat{G}\equiv \widehat{G}_{ab}$, with $a=1,\ldots , 4$,
is a symmetric 4 by 4 metric on
$T^4$ and $ \widehat{C} = \widehat{B} + \dfrac{1}{2} \widehat{A}^{I}
L_{IJ} \widehat{A}^J$ with $\widehat{B}\equiv \widehat{B}_{(0)ab}$.
For each $I$, $\widetilde{A}^I$ is a 4-vector whose components are
$\widetilde{A}^I_{(0)a}$. $L_{IJ}$ is the invariant metric of $O(4, 20)$.
Due to the second identification, the field strengths $\widetilde{F}_{(2)}$, $\widetilde{F}_{(1)m}$ and
$\widetilde{F}_{(0)mn}$ are identified without any change in their expressions with the field strengths
$\widehat{F}_{(2)}$, $\widehat{F}_{(1)m}$ and $\widehat{F}_{(0)mn}$, respectively.

Substituting expressions (\ref{4d-duality}) back into (\ref{IIA-4g}) and (\ref{IIA-4}),
making the identifications
(\ref{MandA}), and changing the sign of the dilaton $\widetilde{\phi} \rightarrow -\widehat{\phi}$,
one obtains the dual Lagrangian, which is
\begin{eqnarray}
\mathcal{L}^{Het}_{4} &=& \frac12 \widehat{R} *1 +\frac18
D\widehat{G}_{mn} \wedge * D\widehat{G}^{mn}  -
D\widehat{\phi}\wedge * D\widehat{\phi} \label{Het-4} \\
& & -\frac14 e^{-2\widehat{\phi}}\widehat{G}_{mn}\mathcal{F}_{(2)}^m \wedge
* \mathcal{F}_{(2)}^{n} -\frac12 e^{2\widehat{\phi}} \lambda_m
\widehat{G}^{mn} \lambda_n \nonumber \\
& & -e^{-\widehat{\phi}} \left[ \dfrac{1}{2}\widehat{H}_{(3)} \wedge * \widehat{H}_{(3)}
+\dfrac{1}{2}\widehat{H}_{(2)m}\widehat{G}^{mn} \wedge * \widehat{H}_{(2)n} \right.
+\dfrac{1}{2}\widehat{H}_{(1)mn} \widehat{G}^{mp}\widehat{G}^{nq}
\wedge * \widehat{H}_{(1)pq} \nonumber \\
& & \left. +\dfrac{1}{2} \widehat{M}_{IJ} \widehat{F}_{(2)}^I \wedge * \widehat{F}_{(2)}^J
+\dfrac{1}{2} \widehat{M}_{IJ} \widehat{F}_{(1)m}^I \widehat{G}^{mn} \wedge *
\widehat{F}_{(1)n}^J +\dfrac{1}{2} \widehat{M}_{IJ} \widehat{F}_{(0)mn}^I \widehat{G}^{mp}
\widehat{G}^{nq} \wedge * \widehat{F}_{(0)pq}^J \right] . \nonumber
\end{eqnarray}

The duality relation between the Lagrangians
(\ref{IIA-4g}-\ref{IIA-4}) and (\ref{Het-4}) is of the S-duality
type, because it changes the sign of the dilaton. Since the string
coupling constant is related to the dilaton with the relation
$g=\exp \phi$, dilaton's sign change corresponds to going from
strong coupling to weak coupling or vice versa.

Now we show that the Lagrangian (\ref{Het-4})  can be obtained
from the six-dimensional Heterotic supergravity Lagrangian through
a twisted reduction on $T^2$. The bosonic sector of Heterotic
supergravity in ten dimensions consists of a scalar dilaton, a
two-form NS-NS potential and gauge bosons $A_{(1)}^a$. It is often
assumed that these vectors take values in the Lie algebra of
$U(1)^{16}$, which is the Cartan subalgebra of either Heterotic
string theory gauge groups, $E_8 \times E_8$ or $Spin(32)/Z_2$.
The six-dimensional Heterotic supergravity Lagrangian is obtained
by the standard Kaluza--Klein reduction of the ten-dimensional
Heterotic supergravity on $T^4$. The details of this reduction can
be found, for example, in \cite{ms}. Like the Type IIA theory in
six dimensions, the six-dimensional Heterotic Lagrangian has rigid
$O(4,20)$ symmetry. Combining fields into multiplets of $O(4,20)$
one can write the Lagrangian in a manifestly $O(4,20)$ invariant
way as
\begin{eqnarray}
\mathcal{L}^{Het}_{6} = e^{-\widehat{\phi}} \left( R * 1 - d\phi \wedge * d\phi +
\dfrac{1}{4} d\widehat{M}_{IJ} \wedge * d\widehat{M}^{IJ}   -\dfrac{1}{2}H_{(3)} \wedge
* H_{(3)} -\dfrac{1}{2} \widehat{M}_{IJ} F_{(2)}^I \wedge * F_{(2)}^J \right) , \nonumber \\ \label{Het-6}
\end{eqnarray}
where $M_{IJ}$  is the $O(4,20)/O(4) \times O(20)$ scalar matrix,
$H_{(3)} = dB_{(2)} - \dfrac{1}{2} L_{IJ} A_{(1)}^I \wedge dA_{(1)}^J$,
and finally $F_{(2)}^I = dA_{(1)}^I$ with $ A_{(1)}^I = ( A_{(1)}^a , B_{(1)  a} , A_{(1)}^A )$.
We can again utilize the $SL(2,R)$ scaling symmetry of $T^2$ to write a twisted reduction
ansatz as\footnote{Note that sign of $\lambda$ in (\ref{SS}) and (\ref{SS2})
have been reversed in each expression except the ones in $A_{(1)}^I$ and $F_{(2)}^I$.}
\begin{eqnarray}
\phi(x, y) & = & \widehat{\phi}(x) + 2 \lambda_m y^m, \ \ \ m=1,2 \  \nonumber \\
G(x,y) & = & e^{\lambda_m y^m}
\left( \widehat{G} + \widehat{G}_{mn}\eta^m \otimes \eta^n \right) , \nonumber \\
B_{(2)}(x,y) & = & e^{\lambda_m y^m} \left( \widehat{B}_{(2)} +
\widehat{B}_{(1)m} \wedge \eta^m
+ \frac12 \widehat{B}_{(0)mn}\eta^m \wedge \eta^n \right) , \nonumber \\
A_{(1)}^I (x,y) & = & e^{1/2 \lambda_m y^m} \left(
\widehat{A}_{(1)}^I + \widehat{A}_{(0)m}^I \wedge \eta^m  \right)
. \label{SSHet}
\end{eqnarray}
This reduction ansatz for the fields involved are the same as
given in (\ref{SS}), except that the metric and three-form field
strength scale differently under the $SL(2,R)$ scaling symmetry.
After inserting the reduction ansatz (\ref{SSHet}) into the
Lagrangian (\ref{Het-6}) we obtain in four dimensions an effective
theory, whose Lagrangian is exactly the same as the dual
Lagrangian given above in (\ref{Het-4}).

%%%%%%%%%%%%%%%%%%%%%%%%%%%%

\section{More on the Massive Duality}

%%%%%%%%%%%%%%%%%%%%%%%%%%%%

In this section, we examine the duality found in
(\ref{4d-duality}) further. We will see how it implies that the
2-form field on the IIA side, which becomes massive by "eating"
one of the 1-form fields (more precisely a linear combination of
two 1-form fields) is dual to a massive 1-form field on the
heterotic side, which  acquires its mass by absorbing the degree
of freedom of  the scalar field. Similarly, the remaining 1-form
field on the IIA side becomes massive by eating the scalar field
and is dual to the 2-form field on the heterotic side, which also
becomes massive due to its St\"{u}ckelberg coupling with the
remaining 1-form field.

It is possible that some fields acquire masses by ``eating``
others due to the St\"{u}ckelberg type couplings between various
fields in equations (\ref{IIA-F}) and (\ref{IIA-H}). Before
explaining how this mechanism works, let us  count the number of
physical degrees of freedom to see in advance how many fields we
expect to become massive in the process. Recall that a massless
$p$-form field in $d$ dimensions has $$C(d-2, p) =
\left(\begin{array}{c}
  d-2 \\
  p \\
\end{array}\right) = \dfrac{(d-2)!}{p! \ (d-2-p)!}$$ \noindent
number of degrees of freedom, whereas the number of physical
degrees of freedom of a massive $p$-form field in $d$ dimensions
is
$$C(d-1, p) = \left(\begin{array}{c}
  d-1 \\
  p \\
\end{array}\right) = \dfrac{(d-1)!}{p! \ (d-1-p)!}. $$
\noindent Then in six dimensions a massless 2-form field has 6
degrees of freedom. If we reduced this 2-form to 4 dimensions with
an ordinary Kaluza-Klein ansatz without the twist, we would obtain
one 2-form field, two 1-form fields and a scalar field, all of which
are massless with a total of $1 + 2 \times 2 + 1 = 6$ degrees of
freedom. However, the twist gives rise to St\"{u}ckelberg type
couplings as above, and upon examination one sees that the 2-form
field eats one of the 1-form fields in 4 dimensions, whereas the
other 1-form field eats the scalar field due to these coupling, as
a result of which we end up with 3+3 = 6 degrees of freedom again.
This will be possible by going to an appropriate gauge as we
explain shortly. Note that a gauge can also be chosen such that
the 1-form fields $A_{(1)}^I$ coming from the reduction of each
$\widehat{A}_{(1)}^I$ eats one of the scalar fields $A_{(0)m}^I$.
However, we prefer not to perform this gauge transformation and
content ourselves with showing the mass gaining mechanism only for
the 2-form field $B_{(2)}$. This is what we will essentially need
when we discuss the duality between the two massive theories
arising from the twisted heterotic and IIA reductions.

In order to explain how the mechanism works, let us first write
down the gauge transformations of the relevant fields in 4
dimensions.
\begin{eqnarray}
\delta B_{(2)} & = & D \Lambda_{(1)} + \Lambda_{(0)1} {\mathcal{F}}^1
+ \Lambda_{(0)2} {\mathcal{F}}^2  \nonumber \\
\delta B_{(1)1} & = & D \Lambda_{(0)1} - \lambda_1 \Lambda_{(1)} \nonumber \\
\delta B_{(1)2} & = & D \Lambda_{(0)2} - \lambda_2 \Lambda_{(1)} \nonumber \\
\delta B_{(0)} & = & -\lambda_1 \Lambda_{(0)2} + \lambda_2 \Lambda_{(0)1}
\end{eqnarray}
\noindent %Here the terms that have been neglected are the terms
%due to the couplings to the fields $A_{(1)}^I$ and $A_{(0)m}^I$
%and will not be relevant for us.
It will be useful to define
$$\overline{B}_{(1)n} = \Omega^m_{\ \ n} B_{(1)m} \ \ \ \ \ \rm{and} \ \ \ \
\ \overline{\Lambda}_{(0)n} = \Omega^m_{\ \ n} \Lambda_{(0)m},$$ \noindent
where
\begin{equation}
\Omega = \frac{1}{\sqrt{\lambda_1^2 +\lambda_2^2}}
\left(\begin{array}{cc}
  \lambda_2 & -\lambda_1 \\
  \lambda_1 & \lambda_2 \\
\end{array}\right).
\end{equation}
\noindent Then we will have
\begin{eqnarray} \delta
\overline{B}_{(1)1} &=& D \overline{\Lambda}_{(0)1} \nonumber \\
\delta \overline{B}_{(1)2} &=& D \overline{\Lambda}_{(0)2} -
\sqrt{\lambda_1^2 + \lambda_2^2} \Lambda_{(1)} \nonumber \\
\delta B_{(0)} & = & \sqrt{\lambda_1^2 + \lambda_2^2} \overline{\Lambda}_{(0)1}
\end{eqnarray}

\noindent Using the last line in the above equation, we can go to
a gauge in which we can set $B_{(0)} = 0$ and the field strengths
in equation (\ref{IIA-H}) become
\begin{eqnarray}
\overline{H}_{(2)1} & = & D \overline{B}_{(1)1} \nonumber \\
\overline{H}_{(2)2} & = & D \overline{B}_{(1)2}
+ \sqrt{\lambda_1^2 + \lambda_2^2} B_{(2)} \nonumber \\
H_{(1)} &=& \overline{B}_{(1)1}.
\end{eqnarray}
\noindent Here $ \overline{B}_{(1)n} = \Omega^m_{\ \ n} B_{(1)m}$ and
$\overline{H}_{(2)n} = \Omega^m_{\ \ n} H_{(2)m}$.
On the other hand, using the gauge invariance
$$ \delta \overline{B}_{(1)2} = -\sqrt{\lambda_1^2 + \lambda_2^2}
\Lambda_{(1)}, \quad {\rm and} \quad \delta B_{(2)} = D \Lambda_{(1)} $$
we can perform the gauge transformation
\begin{equation}
B_{(2)} \longrightarrow B_{(2)} - \dfrac{1}{\sqrt{\lambda_1^2 + \lambda_2^2} } \ D\overline{B}_{(1)2},
\end{equation}
\noindent as a result of which $\overline{B}_{(1)2}$ disappears and $H_{(3)}$ becomes
\begin{equation}
H_{(3)} = DB_{(2)} + \dfrac{1}{\sqrt{\lambda_1^2 + \lambda_2^2}}
(\lambda_1 \overline{B}_{(1)1} {\mathcal{F}}^2 - \lambda_2 \overline{B}_{(1)1} {\mathcal{F}}^1).
\end{equation}

\noindent In summary we will have
\begin{eqnarray}
H_{(3)} &=& DB_{(2)} - \overline{B}_{(1)1} \wedge \overline{{\mathcal{F}}}^1  \nonumber \\
\overline{H}_{(2)1} & = & D \overline{B}_{(1)1} \nonumber \\
\overline{H}_{(2)2} & = &  \sqrt{\lambda_1^2 + \lambda_2^2} B_{(2)} \nonumber \\
H_{(1)} &=& \overline{B}_{(1)1}.
\end{eqnarray}

\noindent Here we defined $\overline{\mathcal{F}}^{m} =
\Omega^m_{\ \ n} {\mathcal{F}}^{n}$ Hence we see that the scalar
field $B_{(0)}$ and the 1-form field $\overline{B}_{(1)2}$ have
been eaten by the 1-form field $\overline{B}_{(1)1}$ and the the
2-form field $B_{(2)}$, respectively. Then, after these special
choices of gauges, the four-dimensional duality relations
(\ref{4d-duality}) become
\begin{eqnarray}
\widehat{D}\widehat{B}_{(2)} - \dfrac{1}{\sqrt{\lambda_1^2 + \lambda_2^2}}
\overline{\widehat{B}}_{(1)1} \wedge {\mathcal{\overline{F}}}^1
-\dfrac{1}{2}L_{IJ}A_{(1)}^I \wedge F_{(2)}^J
& = & e^{-\widetilde{\phi}} *  \overline{\widetilde{B}}_{(1)1},  \\
\sqrt{\lambda_1^2 + \lambda_2^2} \widehat{B}_{(2)}
- L_{IJ} \overline{A}_{(0)2}^I \wedge F_{(2)}^J
&=& - e^{-\widetilde{\phi}} * D\overline{\widetilde{B}}_{(1) 1}, \nonumber \\
 \widehat{D}\overline{\widehat{B}}_{(1)1} - L_{IJ} \overline{A}_{(0)1}^I \wedge F_{(2)}^J
 & = & e^{-\widetilde{\phi}} * \sqrt{\lambda_1^2 + \lambda_2^2} \widetilde{B}_{(2)}, \nonumber \\
\overline{\widehat{B}}_{(1)1} + L_{IJ} A_{(0)1}^I \wedge \widehat{D}A_{(0)2}^J +
\dfrac{1}{2} L_{IJ} \overline{A}_{(0)1}^I \wedge A_{(1)}^J
&=& e^{-\widetilde{\phi}} * (\widetilde{D}\widetilde{B}_{(2)}
- \dfrac{1}{\sqrt{\lambda_1^2 + \lambda_2^2}} \overline{\widetilde{B}}_{(1)1}
\wedge \overline{\mathcal{F}}^1). \nonumber
\end{eqnarray}
\noindent Here $\overline{A}_{(0)n}^I = \Omega^m_{\ \ n}
A_{(0)m}^I$, and fields with $\, \widehat{}\, $ denote the
Heterotic fields and fields with $\, \widetilde{}\, $ denote the
Type IIA fields.  So we see that the duality relation between the
two-form fields in six dimension imply a massive duality relation
between the massive one-form field $\overline{B}_{(1)1}$ and the
massive two-form field $B_{(2)}$ in four dimensions. Note that
both these fields have three degrees of freedom in four
dimensions. This is an illustration in four dimensions of the
general duality between massive $p$-forms and massive
$(d-p-1)$-forms in $d$ dimensions \cite{tpn,q,Bergshoeff:1997ak}.

%%%%%%%%%%%%%%%%%%%%%%%%%%%%

\section{Relation with the gauged $N=4$ supergravity in $D=4$}

%%%%%%%%%%%%%%%%%%%%%%%%%%%%
In this section, we show that the 4 dimensional Lagrangian we
obtained in the previous sections can be put in the form of a $N =
4$ gauged supergravity Lagrangian, whose most general form was
found by Sch\"{o}n and Weidner \cite{sw}. This part of our work is
an extension of the work of \cite{dpp}, where they reduce the
NS-NS sector of the heterotic theory with a twisted ansatz
utilizing the same scaling symmetry and show that the resulting
gauged supergravity is of the Sch\"{o}n-Weidner type with
nontrivial  $SL(2)$ gaugings. Here we also include the vectors
$A^I$ and make the comparison for this more general case.

The $N=4$ supergravity coupled to $n$ vector multiplets has the
global on-shell symmetry  $SL(2,R)\times O(6,6+n)$. The bosonic
sector of the pure $N=4$ supergravity contains the graviton, six
vectors and two scalars, whereas each vector multiplet contains a
vector and six real scalars. The scalar fields of the theory
constitute the coset space $SL(2,R)/SO(2) \times
O(6,6+n)/O(6)\times O(6+n)$.

The gauged $N=4$ supergravities are obtained by gauging a subalgebra of the global symmetry
$SL(2,R)\times O(6,6+n)$. The generators of this subalgebra are the linear combinations of
$SL(2,R)$ and $O(6,6+n)$ generators and the coefficients of
this linear combinations are the components of the
embedding tensor \cite{sw,dst1,dst2,dst3,dst4,dst5,dst6,t,s,w}.
However, there are several requirements which stem form the facts that the
commutator of the generators of subalgebra
should produce an adjoint action, the Jacobi identity for the subalgebra should be satisfied, and the
supersymmetry of the theory should be preserved. Then one obtains a number of constraints that the
components of the embedding tensor have to satisfy \cite{sw}. The components of the embedding
tensor are group valued and usually denoted by $\xi_{\alpha M}$ and $f_{\alpha MNP}=f_{\alpha
[MNP]}$. Here $\alpha = +,-$ is the $SL(2,R)$ index and $M={ i, i^{\prime}, A}$ with $i=1, ... ,
6;\, i^{\prime}=1, ..., 6;$ and $A=1, ... , n$ is the $O(6,6+n)$ index.

The bosonic part of the gauged $N=4$ supergravity action can be written as the sum of a kinetic
term, a topological term and a scalar potential. The kinetic term has the form
\begin{eqnarray}
e^{-1}\mathcal{L}_{kin} & = & \frac12 R*1 +\frac{1}{16}
(D\mathcal{M}_{MN})\wedge *(D\mathcal{M}^{MN})
+\frac18 (DM_{\alpha\beta})\wedge *(DM^{\alpha\beta}) \nonumber \\
& & -\frac14 e^{-2\phi}\mathcal{M}_{MN} H_{(2)}^{M+}\wedge *H_{(2)}^{N+}
+ \frac18 a \eta_{MN} H_{(2)}^{M+}\wedge H_{(2)}^{N+}\ , \label{L-kin}
\end{eqnarray}
the topological term has the form
\begin{eqnarray}
e^{-1}\mathcal{L}_{top} & = & -\frac g2 \left[ \xi_{+M} \eta_{NP}
A_{(1)}^{M-}\wedge A_{(1)}^{N+}\wedge dA_{(1)}^{P+}
-(\hat{f}_{-MNP}+2\xi_{-N} \eta_{MP})
A_{(1)}^{M-}\wedge A_{(1)}^{N+}\wedge dA_{(1)}^{P-} \right. \nonumber \\
& & -\frac g4 \hat{f}_{\alpha MNR} \hat{f}_{\beta PQ}\mbox{}^R
A_{(1)}^{M\alpha}\wedge A_{(1)}^{N+}\wedge A_{(1)}^{P\beta}\wedge A_{(1)}^{Q-}
+\frac g{16} \Theta_{+MNP} \Theta_{-}\mbox{}^M \mbox{}_{QR}
B_{(2)}^{NP}\wedge B_{(2)}^{QR} \nonumber \\
& & \left. -\frac14 ( \Theta_{-MNP} B_{(2)}^{NP} + \xi_{-M} B_{(2)}^{+-}
+ \xi_{+M} B_{(2)}^{++} )\wedge ( 2dA_{(1)}^{M-}
- g\hat{f}_{\alpha QR}\mbox{}^M A_{(1)}^{Q\alpha}\wedge A_{(1)}^{R-} ) \right]\ , \nonumber \\
\label{L-top}
\end{eqnarray}
and the scalar potential term has the form
\begin{eqnarray}
e^{-1}\mathcal{L}_{pot} & = & -\frac{g^2}{16} \left[ f_{\alpha MNP} f_{\beta QRS} M^{\alpha\beta}
\left( \frac13 \mathcal{M}^{MQ}\mathcal{M}^{NR}\mathcal{M}^{PS}
+ \left(\frac23 \eta^{MQ} - \mathcal{M}^{MQ}\right)
\eta^{NR}\eta^{PS} \right) \right. \nonumber \\
& & \left. -\frac49 f_{\alpha MNP} f_{\beta QRS} \epsilon^{\alpha\beta}M^{MNPQRS}
+ 3\xi_{\alpha}^{M} \xi_{\beta}^{N} M^{\alpha\beta}\mathcal{M}_{MN} \right]\ . \label{L-pot}
\end{eqnarray}

Now let us explain the terms that appear in the above action:
$\eta^{MN}$ is the $O(6,6+n)$ metric, which can be written in
blocks as
\begin{equation}
\eta^{MN} = \left(
\begin{array}{ccc}
 0\ & 1\ & 0  \\
 1\ & 0\ & 0  \\
 0\ & 0\ & L  \\
\end{array}
\right)\ ,
\end{equation}
where $L^{IJ}$ is the $O(4,4+n)$ metric. $\mathcal{M}_{MN}$ is a
symmetric positive definite scalar matrix that parametrize the
coset manifold $O(6,6+n)/O(6)\times O(6+n)$, likewise
$M_{\alpha\beta}$ is a symmetric positive definite matrix that
parametrizes the $SL(2,R)/SO(2)$ coset space. A suitable choice
for $M_{\alpha\beta}$ is
\begin{equation}
\label{Mab}
M_{\alpha\beta}=\frac{1}{Im(\tau )}\left(
\begin{array}{cc}
|\tau |^2 & Re(\tau ) \\
Re(\tau ) & 1
\end{array} \right) .
\end{equation}
where  $\tau =a + ie^{-2\phi }$, $a$ is the axion and $\phi$ is
the dilaton field. \noindent The  $M_{MNPQRS}$ that appear in the
scalar potential term of the Lagrangian is a scalar dependent,
completely antisymmetric tensor, which is also defined in terms of
$O(6,6+n)/O(6)\times O(6+n)$ coset representatives \cite{sw}.

The components of the embedding tensor, $\xi_{\alpha M}$ and $f_{\alpha MNP}=f_{\alpha
[MNP]}$, frequently appear in the following combinations:
\begin{eqnarray}
\Theta_{\alpha MNP} & = & f_{\alpha MNP} - \xi_{\alpha [N}\eta_{P]M}\ , \nonumber \\
\hat{f}_{\alpha MNP} & = & f_{\alpha MNP} - \xi_{\alpha [M}\eta_{P]N}
- \frac32 \xi_{\alpha N}\eta_{MP}\ . \label{f-hat}
\end{eqnarray}
The gauge coupling constant $g$ will later be taken as $g=1$. The covariant derivative is defined as
\begin{equation}
\label{Dmu}
D = \nabla -gA_{(1)}^{M\alpha}\Theta_{\alpha M}\mbox{}^{NP}t_{NP}
+gA_{(1)}^{M(\alpha}\epsilon^{\beta)\gamma}\xi_{\gamma M}t_{\alpha\beta}\ ,
\end{equation}
where $t_{NP}$ and $t_{\alpha\beta}$ are generators of $O(6,6+n)$ and $SL(2,R)$, respectively
and $\nabla$ contains the spin connection. $D$ acts on objects in an arbitrary
representation of the global symmetry group.

$H_{(2)}^{M+}$ are the covariant field strengths of electric fields with forms given as
\cite{sw}
\begin{eqnarray}
H_{(2)}^{M+} &=& F_{(2)}^{M+} +\frac g2 \Theta_{-}\mbox{}^M \mbox{}_{NP}B_{(2)}^{NP}
+\frac g2 \xi_{+}\mbox{}^M B_{(2)}^{++} +\frac g2 \xi_{-}\mbox{}^M B_{(2)}^{+-} \label{Hmn+} , \\
 F_{(2)}^{M+} &=& dA_{(1)}^{M+} -g\hat{f}_{\alpha NP}\mbox{}^M
A_{(1)}^{N\alpha}\wedge A_{(1)}\mbox{}^{P+} \label{Fmn+} .
\end{eqnarray}
The covariant field strengths of magnetic fields, $H_{(2)}^{M-}$,
are defined similarly by interchanging all $-$ and $+$ indices in
the above expression. Note that there is no kinetic term for the
magnetic fields $H_{(2)}^{M-}$.

To match the the four-dimensional Heterotic string Lagrangian (\ref{Het-4}) with the four-dimensional
gauged supergravity Lagrangian (\ref{L-kin}--\ref{L-pot})
we have to make several field definitions and define a
$O(6,22)/O(6) \times O(22)$ valued scalar matrix which could be matched with $\mathcal{M}^{MN}$.
We find that after following field definitions\footnote{Note that $\widehat{C}_{(0)mn}$ is
not antisymmetric in its indices. In fact, in matrix notation we
have $ \widehat{C} = \widehat{B} + \dfrac{1}{2} \widehat{A}^T L \widehat{A}$
so that $ \widehat{C} + \widehat{C}^T = \widehat{A}^T L \widehat{A}$ rather than 0.}
\begin{eqnarray}
\widehat{C}_{(2)} &=& \widehat{B}_{(2)} + \frac12 \widehat{B}_{(1)m}\wedge \mathcal{A}_{(1)}^m
+ \frac14 L_{IJ} \widehat{A}_{(1)}^I \wedge \widehat{A}_{(0) m}^J \wedge  {\mathcal{A}}_{(1)}^m\\
\widehat{C}_{(1)m} &=& \widehat{B}_{(1)m}
+ \dfrac{1}{2} L_{IJ} \widehat{A}_{(1)}^I \wedge \widehat{A}_{(0)m}^J \label{C1m} \\
\widehat{C}_{(0)mn} &=& \widehat{B}_{(0)mn} + \dfrac{1}{2} L_{IJ}
\widehat{A}_{(0)m}^I \wedge \widehat{A}_{(0)n}^J \label{C0mn}
\end{eqnarray}
the field strengths $\widehat{H}_{(3)}$, $\widehat{H}_{(2)m}$ and $\widehat{H}_{(1)mn}$ can be written as
\begin{eqnarray}
\widehat{H}_{(3)}  & = & \widehat{D}\widehat{C}_{(2)} - \frac12 \eta_{MN}
\widehat{\mathcal{A}}_{(1)}^M \wedge \widehat{\mathcal{F}}_{(2)}^N
- \frac14 \lambda_m {\mathcal{A}}_{(1)}^m \wedge
\widehat{C}_{(1) p} \wedge {\mathcal{A}}_{(1)}^p \label{Het-H3} \\
\widehat{H}_{(2)m} & = & \widehat{D}\widehat{C}_{(1)m} + \lambda_m
\widehat{C}_{(2)} - \widehat{C}_{(0)mn} \wedge
{\mathcal{F}}_{(2)}^n - L_{IJ} \widehat{A}_{(0)m}^I \wedge
D\widehat{A}_{(1)}^J \label{Het-H2} \\
\widehat{H}_{(1)mn} & = & \widehat{D}\widehat{C}_{(0)mn} - L_{IJ}
\widehat{A}_{(0) m}^I \wedge D\widehat{A}_{(0) n}^J , \label{Het-H1}
\end{eqnarray}
where $D\widehat{A}_{(p)}^I$ are as in (\ref{IIA-F}), and we define
$ \widehat{\mathcal{A}}_{(1)}^M = (\mathcal{A}_{(1)}^m ,
\widehat{C}_{(1)  m} , \widehat{A}_{(1)}^I )$ and their field strengths,
\begin{equation}
\widehat{\mathcal{F}}_{(2)}^M = (\mathcal{F}_{(2)}^m ,
\widehat{D}\widehat{C}_{(1) m}+ \lambda_m \widehat{C}_{(2)} , D\widehat{A}_{(1)}^I ) .
\label{cal-F}
\end{equation}
On the other hand the covariant derivatives have the following forms:
\begin{eqnarray}
\widehat{D}\widehat{C}_{(2)} &=& d\widehat{C}_{(2)}
- \lambda_r \widehat{C}_{(2)} \wedge {\mathcal{A}}_{(1)}^r \\
\widehat{D}\widehat{C}_{(1)m} &=& d\widehat{C}_{(1)m}
+ \lambda_r \widehat{C}_{(1)m}\wedge {\mathcal{A}}_{(1)}^r
-\frac12 \lambda_m \widehat{C}_{(1) p} \wedge {\mathcal{A}}_{(1)}^p  \\
\widehat{D}\widehat{C}_{(0)mn} &=& d\widehat{C}_{(0)mn} -
\lambda_r \widehat{C}_{(0)mn} \wedge {\mathcal{A}}_{(1)}^r - 2
\lambda_{[m}\widehat{C}_{(1)n]}\ . \label{covdevC}
\end{eqnarray}
We also define the $O(6,22)/O(6) \times O(22)$ valued scalar matrix as
\begin{equation}
\mathcal{\widehat{N}}^{MN} = \left(\begin{array}{ccc}
  \widehat{G} + \widehat{C}^T \widehat{G}^{-1} \widehat{C}
  + \widehat{A}^{T} \widehat{M} \widehat{A}^J
  & -\widehat{C}^T \widehat{G}^{-1} & \widehat{C}^T \widehat{G}^{-1}L \widehat{A}
  + \widehat{A}^{T} \widehat{M} \\
  -\widehat{G}^{-1} \widehat{C} & \widehat{G}^{-1} & -\widehat{G}^{-1} L \widehat{A} \\
  \widehat{A}^T L \widehat{G}^{-1} \widehat{C} + \widehat{M}\widehat{A} &
  -\widehat{A}^T L \widehat{G}^{-1} & \widehat{M} + \widehat{A}^T L \widehat{G}^{-1} L \widehat{A} \\
\end{array}\right). \label{N}
\end{equation}
where $\widehat{G}\equiv \widehat{G}_{mn}$, with $m=1, 2$,
is a symmetric 2 by 2 metric on $T^2$ and $ \widehat{C} = \widehat{B} + \dfrac{1}{2} \widehat{A}^{I}
L_{IJ} \widehat{A}^J$ with $\widehat{B}\equiv \widehat{B}_{(0)mn}$.
For each $I$, $\widetilde{A}^I$ is a 2-vector whose components are
$\widetilde{A}^I_{(0)m}$. $L_{IJ}$ is the invariant metric of $O(4, 20)$, and
$O(4,20)/O(4) \times O(20)$ valued scalar matrix $\widehat{M}^{IJ}$ is given in (\ref{Het-M}).

We can now rewrite the four-dimensional Heterotic Lagrangian (\ref{Het-4}) in a form which is
ready to be compared with the four-dimensional supergravity Lagrangian:
\begin{eqnarray}
\mathcal{L}^{Het}_{4} &=& \frac12 \widehat{R}*1
- (D \widehat{\phi})\wedge *(D \widehat{\phi})
-\frac12 e^{2\widehat{\phi}} \lambda_m \widehat{\mathcal{N}}^{mn} \lambda_n \nonumber \\
& &  + e^{-\widehat{\phi}} \left[
\dfrac{1}{4} D\mathcal{\widehat{N}}_{MN} \wedge * D\mathcal{\widehat{N}}^{MN}
-\dfrac{1}{2}\widehat{H}_{(3)} \wedge * \widehat{H}_{(3)}
-\dfrac{1}{2} \mathcal{\widehat{N}}_{MN}
\widehat{\mathcal{F}}_{(2)}^M \wedge * \widehat{\mathcal{F}}_{(2)}^N \right] . \label{Heter-4}
\end{eqnarray}

Now we need to solve for $\xi_{\alpha M}$ and $f_{\alpha MNP}$ in
order to bring the Lagrangian of the gauged supergravity in four
dimensions (\ref{L-kin}--\ref{L-pot}) to a form that is equivalent
to the four-dimensional Heterotic Lagrangian (\ref{Heter-4}).
However, instead of solving for possible $\xi_{\alpha M}$ and
$f_{\alpha MNP}$ from constraint equations (eq. (2.20) in
\cite{sw}), we  determine them by comparing the field strengths
(\ref{Hmn+}) in four-dimensional gauged supergravity Lagrangian
with the field strengths $\widehat{\mathcal{F}}_{(2)}^M$ in the
Heterotic supergravity Lagrangian. Then it can be shown that the
 solution we find indeed obeys the constraint equations of
\cite{sw}.

In (\ref{Heter-4}) we have only the field strengths of electric fields. Therefore we first set
$\xi_{- M}=0$ and $f_{- MNP}=0$. Now comparing the field strength
$\widehat{\mathcal{F}}_{(2)}^M$ (\ref{cal-F}) with $H_{(2)}^{M+}$ (\ref{Hmn+})
we firstly observe that to have an equivalence we need to identify
$B_{(2)}^{++}$ with $2\widehat{C}_{(2)}$ and set the values of $\xi_{+ M}$ as
\begin{equation}
\xi_{+ M} =  (\xi_{+ m}, \xi_{+ m^{\prime}}, \xi_{+ I}) = (\lambda_{m}, 0, 0) . \label{xi}
\end{equation}
The other observation is about the values of $\hat{f}_{+ MNP}$, which we obtain as
\begin{eqnarray}\label{aybike}
\hat{f}_{+ MNp^{\prime}} - \hat{f}_{+ NMp^{\prime}} & = & 0 \nonumber \\
\hat{f}_{+ m^{\prime}np} - \hat{f}_{+ nm^{\prime}p} & = & \lambda_{p}\eta_{nm^{\prime}}
-2\lambda_{n}\eta_{pm^{\prime}} \nonumber \\
\hat{f}_{+ ImJ} - \hat{f}_{+ mIJ} & = & -\lambda_{m}\eta_{IJ}
\end{eqnarray}
Using the definition (\ref{f-hat}) of $\hat{f}_{+ MNP}$, values of $\xi_{+ M}$ (\ref{xi}) and the
antisymmetry property $f_{+ MNP}=f_{+ [MNP]}$, we can now determine that
\begin{equation}
f_{+ mnp^{\prime}} = -\lambda_{[m}\eta_{n]p^{\prime}} , \label{f}
\end{equation}
with all other components of $f_{+ MNP}$ vanishing. Here,
$\lambda_{m}$ has only two components $\lambda_{1}$ and
$\lambda_{2}$, unlike \cite{dpp}. This is because we put fluxes
only on $T^2$, whereas in \cite{dpp}, a twisted reduction on the
6-torus $T^6$ is considered. Note that the components of the
tensor $f_{+MNP}$ involving the indices $I$ are zero, in spite of
the non-Abelian field strengths of the vector fields
$\widehat{A}^I$. Comparing $D\widehat{A}^I$ with (\ref{Hmn+}), one
finds the last equation in (\ref{aybike}) above, yet the
components $f_{+mIJ}$ are computed to be zero. As a result, our
embedding tensor contains no new nonvanishing components as
compared to the one found in \cite{dpp}. We refer the reader to
\cite{sw} to check that the solution (\ref{f}) satisfies the
constraint equations that the embedding tensor should satisfy.

Plugging in the determined values for embedding tensor components
(\ref{xi}, \ref{f}) and then integrating out the magnetic fields,
$A_{(1)}^{M-}$, from the gauged supergravity Lagrangian
(\ref{L-kin}--\ref{L-pot}) one obtains that the combination of the
kinetic part and the topological part become \cite{dpp}
\begin{eqnarray}
e^{-1}\mathcal{L}_{kin} & = & \frac12 R*1
+\frac{1}{16}(D\mathcal{M}_{MN})\wedge *(D\mathcal{M}^{MN})
- (D\phi)\wedge *(D\phi) \label{L-kin2} \\
& & -\frac14 e^{-2\phi}\mathcal{M}_{MN}
\left( F_{(2)}^{M+} + \frac12 \xi_+^M B_{(2)}^{++} \right)\wedge
*\left( F_{(2)}^{N+} + \frac12 \xi_+^N B_{(2)}^{++} \right) \nonumber \\
& & - \frac18 e^{-4\phi} \left( dB_{(2)}^{++} - \xi_{+M}
A_{(1)}^{M+}\wedge B_{(2)}^{++} - \omega_{(3)} \right) ^2\ , \nonumber
\end{eqnarray}
and the scalar potential term become
\begin{eqnarray}
e^{-1}\mathcal{L}_{pot} & = & -\frac{1}{16} e^{2\phi} \left[ 3 \xi_{+M}\xi_{+N}  \mathcal{M}^{MN}
+ \frac13 f_{+ MNP} f_{+ QRS}
\mathcal{M}^{MQ}\mathcal{M}^{NR}\mathcal{M}^{PS} \right. \nonumber \\
& & \left. \makebox[1.4cm]{} + f_{+ MNP} f_{+ QRS} \left( \frac23 \eta^{MQ} - \mathcal{M}^{MQ}\right)
\eta^{NR}\eta^{PS} \right] \ ,  \label{L-pot2}
\end{eqnarray}
where $\omega_{(3)}= \eta_{MN} F_{(2)}^{M+} \wedge A_{(1)}^{N+} -
\frac12 \lambda_m \mathcal{A}_{(1)}^{m+} \wedge A_{(1)}^{n+}
\wedge A_{(1) n}^+$ and we set $g=1$. Note that $\widehat{H}_{(3)}
= \widehat{D}\widehat{C}_{(2)} - \dfrac{1}{2} \omega_{(3)}$.

We note that identifying $A_{(1)}^{m+}$ with the Kaluza-Klein
gauge fields $\mathcal{A}_{(1)}^m$, $A_{(1)}^{m^{\prime}+}$ with
the field $\widehat{C}_{(1)m}$ (\ref{C1m}), $A_{(1)}^{I+}$ with
the vector fields of six dimensions, and $B_{(2)}^{++}$ with
$2\widehat{C}_{(2)}$ one matches the kinetic terms of gauge fields
in four-dimensional gauged supergravity Lagrangian (\ref{L-kin2})
with the kinetic terms of gauge fields in the four-dimensional
Heterotic string Lagrangian (\ref{Heter-4}). One needs also to
check whether the scalar potential term (\ref{L-pot2}) of the
four-dimensional gauged supergravity action matches with the
scalar potential term in the above Lagrangian after the
identification $\mathcal{\widehat{M}}\equiv 2
\mathcal{\widehat{N}}$. Substituting in the scalar potential term
(\ref{L-pot2}) the matrix form of $\mathcal{\widehat{N}}$ one
finds that scalar potential terms also match. This way we show
that the compactification of heterotic string theory with the
inclusion of the Yang-Mills vectors is equivalent to a gauged
supergravity which is still of the Sch\"{o}n-Weidner type.

%%%%%%%%%%%%%%%%%%%%%%%%%%%%

\section{Conclusions}

%%%%%%%%%%%%%%%%%%%%%%%%%%%%

In this paper, we established a massive S-duality  relation
between the heterotic theory and type IIA theory in 4 dimensions.
Both theories in four dimensions are obtained by a duality-twisted
reduction, which exploits the scaling symmetry of various fields
including the dilaton and  the metric. This type of reduction
ansatz was first used by Derendinger et al. \cite{dpp} for the
reduction of the NS-NS sector of the heterotic theory. Our ansatz
for the reduction of heterotic and type IIA theories assign gauge
coupling of the opposite sign to the NS-NS fields and couplings of
the same sign to the 1-form fields. The massive duality between
the two theories work in the following way. On the one side we
have scalar, vector and 2-form fields ($p-$form fields with
$p=0,1,2$) with various St\"{u}ckelberg type couplings. Such
couplings allow a $p-$form field to become massive by absorbing
the degrees of freedom of a $(p-1)$-form field after a certain
gauge transformation. In the massless case, a $p-$form field is
dual to a $(\tilde{p} = 2 - p)$-form field in 4 dimensions.
Similarly, a $(p-1)$-form field is dual to a $(3-p = \tilde{p} +
1)$-form field. These dual fields also have St\"{u}ckelberg type
couplings among them. As a result, the $(\tilde{p}+1)$-form fields
absorb the degrees of freedom of the $\tilde{p}$-form fields and
hence become massive. This massive $(\tilde{p}+1 = 3 - p)$-form
field is the dual of the massive $p$-form field in the original
theory. The duality between the two theories also changes the sign
of the dilaton, and therefore it is of the S-duality type. So we
see that the usual S-duality between the heterotic and IIA
theories in 4 dimensions survive, even in the presence of (a
certain class of) fluxes.

In the last section of our paper, we also showed that the Lagrangian
for the massive theory we obtain in four dimensions can be put in
the general form of $N=4, D=4, SL(2) \times O(6,22)$ gauged
Lagrangian, found by Sch\"{o}n and Weidner \cite{sw}, where (part of)
the $SL(2)$ group has been gauged. This had already been done by
Derendinger et al. \cite{dpp} for the NS-NS sector of the heterotic
theory. Here, we also add the sector involving the vector fields
coming from the reduction of the Yang-Mills vectors in 10
dimensions, and show that the resulting theory is still of the
same type.

A natural generalization of our work would be to introduce a more
general duality-twisted ansatz, which also gauges the $O(6,22)$
part (and even more interestingly the whole of the $SL(2)$ part)
of the symmetry group in 4 dimensions and explore the faith of
S-duality in this more general case.

Another interesting direction is to analyze if the
string-string-string triality in 4 dimensions \cite{dlm}
continues to hold in the presence of fluxes we consider here. It
would be very interesting to find a duality-twisted ansatz for the
reduction of type IIB theory, which gives in 4 dimensions a
massive theory dual to the two massive theories we have found
here.

%%%%%%%%%%%%%%%%%%%%%%%%%%%%

\acknowledgments

This work is supported by the Turkish Council of
Research and Technology (T\"{U}B\.{I}TAK) through grant number
108T715.

%%%%%%%%%%%%%%%%%%%%%%%%%%%%

\end{document}